\documentclass[draft]{agujournal}
\draftfalse
\usepackage{bm,soul}

\begin{document}

\title{Turbulent kinematic dynamos in ellipsoids\\ driven by mechanical forcing}

\authors{K. Sandeep Reddy\affil{1}, Benjamin Favier\affil{1}, Michael Le Bars\affil{1}}

\affiliation{1}{CNRS, Aix Marseille Univ, Centrale Marseille, IRPHE, Marseille, France}

\correspondingauthor{Favier}{favier@irphe.univ-mrs.fr}

\begin{keypoints}
\item MHD simulations of turbulent flows driven by mechanical forcing in ellipsoids are performed
\item Planetary libration, precession and tides are all capable of dynamo action
\item Our results validate previous and encourage new unconventional scenarii for planetary magnetism

\end{keypoints}

\begin{abstract}
Dynamo action in planetary cores has been extensively studied in the context of convectively-driven flows.
We show in this letter that mechanical forcings, namely tides, libration and precession, are also able to kinematically sustain a magnetic field against ohmic diffusion.
Previous attempts published in the literature focused on the laminar response or considered idealized spherical configurations.
In contrast, we focus here on the developed turbulent regime and we self-consistently solve the magnetohydrodynamic (MHD) equations in an ellipsoidal container.
Our results open new avenues of research in dynamo theory where both convection and mechanical forcing can play a role, independently or simultaneously.
\end{abstract}

\section{Introduction}

The existence of a self-sustained magnetic field by dynamo action constitutes the most obvious signature of the intense fluid dynamics taking place or having taken place in the liquid iron core of terrestrial planets.
It is also a key ingredient for their habitability.
As such, it has been the subject of numerous studies since the seminal work of \citet{larmor1919possible}, and it remains today an active and challenging research domain for theoretical, experimental and numerical approaches \citep[see e.g.,][]{Busse2015239, cardin2015, christensen2015numerical, roberts20158}.
One of the prevailing questions is the source of fluid motions.
On Earth today, it is generally agreed that the dynamo is driven by turbulent flows sustained by convection coming from the combination of thermal and compositional buoyancy, with a major role played by the solidification of the inner core \citep[][]{roberts20158}.
The energy budget is, however, extremely tight with related  unsolved questions regarding the heat flux at the core-mantle boundary (CMB), the age of the inner core, and the presence of stratified layers at the boundaries \citep[see e.g.,][]{labrosse2015thermal}.
The situation is even more complex before the onset of inner core solidification, where dynamo action can hardly be sustained by thermal convection only and necessitates additional effects \citep[][]{badro2016early,o2016powering}.
Besides Earth, the convective dynamo model has been generically applied to other terrestrial bodies, in the context of an often tacitly assumed equivalence between the presence of planetary magnetism and of core convection.
The standard convective dynamo model is however hardly capable of explaining the past or present magnetic field of numerous bodies like the Moon, Ganymede, Mercury, Mars, large asteroids, etc. \citep[][]{Dwyer2011, Lebars2011, sarson1997magnetoconvection, arkani2008tidal, wei2014simplified}. Even if more evolved or specific convective scenarii might provide plausible explanation,  involving for instance thin shell dynamics \citep[e.g.][]{stanley2005thin}, non-homogeneous or time-dependent boundary conditions  \citep[e.g.][]{stanley2005thin, cao2014dynamo, roberts2017effects}, a stably stratified layer \citep[e.g.][]{tian2015magnetic}, iron snow \citep[e.g.][]{christensen2015iron}, this has led some researchers to look for alternative routes for core flows and dynamo.

Following the seminal ideas of \citet{Malkus1963, Malkus:Science1968,Malkus1989}, a huge reservoir of energy is stored in the rotational motion of terrestrial bodies, including their spinning and orbital motions.
If those bodies were rotating with a perfectly constant rotation vector and with a perfectly rigid shape, the fluid in their core, in the absence of convection, would simply follow the rotation of their mantle like a solid body.
However, this is never the case. First, differential rotation might be present in planetary cores, which is capable of sustaining dynamos \citep[e.g.][]{guervilly2010numerical,cao2012saturn}. Then, the rotation of the planetary bodies is always perturbed by gravitational interactions with neighbors. Indeed, the angular velocity of the rotating bodies changes periodically with time, corresponding to longitudinal libration and length of day variation; the direction of the rotational vector changes periodically with time, corresponding to precession and nutation; and the shape of the terrestrial bodies changes periodically with time, corresponding to tidal distortions.
All these perturbations, generically called harmonic or mechanical forcings, are small in amplitude. For example, on Earth, the relative amplitude of the tidal bulge vs. the core radius as well as the relative amplitude of the precession vs. rotation rates are about $10^{-7}$ only.
This smallness of the perturbation amplitude has led to the misinterpretation that the resulting flows are also small \citep[][]{loper1975torque, rochester1975can}.
But beyond the directly forced laminar base flow, harmonic forcings can sustain instabilities via the resonant excitation of the eigenmodes of the rotating (and possibly stratified) bodies, i.e. the so-called (gravito-) inertial waves \citep[][and references therein]{le2015flows}.
Those instabilities take their energy from the huge rotational energy of the system, the weak forcing being only the conveyor: instabilities then potentially drive intense, highly energetic, bulk filling turbulence in the whole core \citep[][]{Kerswell1996}.

Libration, precession and tidal instabilities have been the subject of numerous studies focusing on their threshold and linear growth \citep[][and references therein]{le2015flows}, and more recently on their non-linear saturation \citep[e.g.,][]{barker2013non,lin2015shear, le2017inertial}, combining theoretical, experimental, and numerical approaches.
The relevance of those alternative sources of core turbulence for terrestrial bodies has also been the subject of several studies \citep[e.g.,][]{Seyed2004,Cebron_2012,Grannan:GJI2016, lemasquerier2017libration}, and has given birth to unconventional scenarii to explain past or existing dynamos: e.g., on Io \citep[][]{Kerswell1998}, on Mars \citep[][]{arkani2008tidal}, on the Moon \citep[][]{Dwyer2011, Lebars2011}, on the early Earth \citep[][]{andrault2016deep}.
However, studies of the dynamo capability of the flows resulting from libration, precession and tides, have been up to now sparse and limited to idealized or simplified configurations: i.e. laminar dynamos from the precession base flow \citep[][]{ernst2013}, laminar dynamos for tidal instability with adhoc bulk forcing in a spherical domain \citep[][]{Cebron:AJL2014,vidal2017magnetic}, laminar dynamos in a spheroidal domain for precession and libration instabilities \citep[][]{Wu:GAFD2009, Wu:GAFD2013} \citep[but see also][]{guermond2013remarks}, and turbulent dynamos in a spherical domain for precession \citep[][]{Tilgner:POF2005, tilgner2007kinematic, kida2011turbulent, Lin:POF2016}.
However, relevance to planetary configurations implies considering fully turbulent flows in non-axisymmetric ellipsoidal geometry, accounting for the static and/or dynamic tidal distortions of the CMB.
This is extremely challenging from a numerical point of view. We present here the first realizations of such kinematic tidal, librational, and precessional turbulent dynamos.

\section{Model formulation}

We present in the following only the main features of our numerical model. All technical details and benchmarks can be found in the Supporting Information \citep[][]{Cattaneo:PRL1995,Chan:PEPI2007,Dudley:PRSA1989,Galloway:Nature1992,Iskakov:JCP2004}.

\subsection{Inner fluid Domain}

We consider the motion of an incompressible, electrically-conducting fluid enclosed in an ellipsoidal cavity surrounded by a large insulating domain.
The density $\rho$, kinematic viscosity $\nu$ and magnetic diffusivity $\eta$ of the fluid are assumed constant.
The outer surface of the ellipsoidal fluid cavity is generally defined using Cartesian coordinates $(x,y,z)$ as $(x/a)^2+(y/b)^2+(z/c)^2=1$ where $a$, $b$ and $c$ are constants.
Following \citet{Grannan:GJI2016}, we non-dimensionalize our problem using the mean equatorial radius $R=\sqrt{(a^2+b^2)/2}$, leading to the following dimensionless definition of the ellipsoidal boundary
\begin{linenomath*}
\begin{equation}
\label{Eq:Ellip_f}
\frac{x^2}{1+\beta} + \frac{y^2}{1-\beta} +\frac{z^2}{c_*^2} = 1 \ ,
\end{equation}
\end{linenomath*}
with the equatorial ellipticity $\beta=(a^2-b^2)/(a^2+b^2)$ and the axial length $c_*=c/R$.

\begin{figure}[ht!]
\centerline{\includegraphics[width=\linewidth]{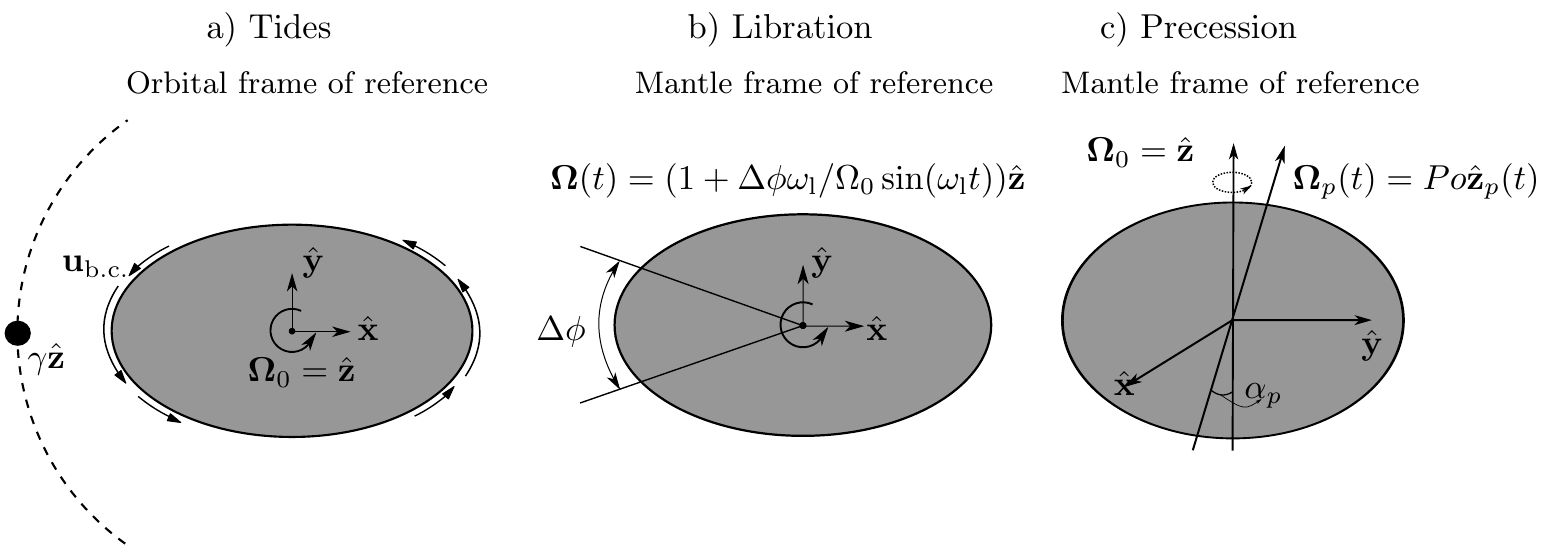}}
\caption{Schematic of the three configurations considered here. a) Tides: in the orbital frame rotating with the ellipsoidal deformation (i.e. where the tidal bulge is fixed), the fluid is entrained by a tangential velocity at the ellipsoidal boundary. b) Libration: the mantle is rotating with the sum of a constant rotation rate $\Omega_0$ and a periodic modulation, driving the fluid through Coriolis and Poincar\'e forces. c) Precession: the mantle is rotating around two different axes, $\hat{\mathbf{z}}$ and $\hat{\mathbf{z}}_p$, at two different rates given by the Poincar\'e number $Po=\Omega_p/\Omega_0$.}
\label{Fig:schema}
\end{figure}

We work in a non-inertial frame of reference rotating with a dimensionless angular velocity $\bm{\Omega}(t)$, normalized by an average rotation rate $\Omega_0$.
The definition of $\bm{\Omega}(t)$ depends on the case of interest and will be detailed in the following sections.
In all cases, we choose to work in the frame in which the mantle shape is fixed in space and time.
Scaling time with ${\Omega^{-1}_0}$, space with $R$, and vector potential field with $R^2\Omega_0\sqrt{\rho \mu_0}$, where $\mu_0$ is the vacuum magnetic permeability, the nondimensionalised MHD equations in the fluid domain, where the dynamo action takes place, are 
\begin{eqnarray}
\frac{\partial\mathbf{u}}{\partial t} + (\mathbf{u}\cdot\nabla)\mathbf{u}
 &=& -\nabla \Pi +  E\nabla^2 \mathbf{u}+\mathbf{F}_i \label{Eq:NS} \\
\frac{\partial\mathbf{A}}{\partial t}  &=&  {\mathbf{ u \times(\nabla \times A})} + \frac{E}{Pm}\nabla^2 \mathbf{A} \label{Eq:VP_f} \\
\nabla \cdot \mathbf{u} &=& 0 \label{Eq:incom} \\
\nabla \times \mathbf{A} &=& \mathbf{B}
\end{eqnarray}
where $\mathbf{u}$, $\mathbf{A}$, and $\mathbf{B}$ are the non-dimensional velocity, vector potential and magnetic field, respectively.
$\Pi$ is the modified pressure including the centrifugal acceleration and the gravitational potential.
We have neglected the Lorentz force in equation~(\ref{Eq:NS}), focusing on the kinematic problem.
This general problem involves two dimensionless parameters: the magnetic Prandtl number $Pm=\nu/\eta$ and the Ekman number $E=\nu/(\Omega_0 R^2)$.
We use the vector potential approach \citep{Matsui:EPS2004,Matsui:IJCFD2004,Cebron:GAFD2012} for numerical reasons (see details in the Supporting Information), thus exactly satisfying the divergence-free condition for the magnetic field.
$\mathbf{F}_i$ stands for the fictitious Coriolis and Poincar{\'e} forces, generally given by
\begin{linenomath*}
\begin{equation}
\mathbf{F}_i=-\underbrace{2\boldsymbol{\Omega}(t)\times \mathbf{u}}_\mathrm{Coriolis}-\underbrace{\frac{\partial \boldsymbol{\Omega}(t)}{\partial t} \times \mathbf{r}}_{\textrm{\scriptsize Poincar\'e}} \ , 
\end{equation}
\end{linenomath*}
where $\mathbf{r}$ is the position vector. We now detail the three cases of interest and their respective inertial forces and frame of reference (see also Figure~\ref{Fig:schema}).

\subsubsection{Tides}

We consider an idealized tidally-driven flow as detailed in \cite{CEBRON2010} and \citet{Grannan:GJI2016}.
The frame of reference is attached to the elliptical distortion which is rotating at a constant rate $\Omega_{\mathrm{orb}}\mathbf{\hat{z}}$, hereby named as orbital frame.
The dimensionless inertial force $\mathbf{F}_i$ in equation (\ref{Eq:NS}) is therefore just a Coriolis force given by
\begin{linenomath*}
\begin{equation}
\mathbf{F}_i = -2 \gamma\mathbf{\hat{z}}\times\mathbf{u},
\end{equation}
\end{linenomath*}
where $\gamma = \Omega_\mathrm{orb}/{\Omega}_0$ is the ratio between the rotation rate of the ellipsoidal deformation and the absolute spin rate ${\Omega}_0$ of the mantle. 
In this orbital frame, the core--mantle boundary rotates at a differential rotation rate $\Omega_0-\Omega_{\mathrm{orb}}$.  
Following \citet{CEBRON2010}, a no-slip condition is thus imposed at the inner domain boundary, given by
\begin{linenomath*}
\begin{equation}
\label{eq:uc_tide}
\mathbf{u}_\mathrm{b.c.} = (1-\gamma)\sqrt{1-\frac{z^2}{c_*^2}}\hat{\boldsymbol{\tau}} \ , 
\end{equation}
\end{linenomath*}
where $\hat{\boldsymbol{\tau}}$ is the normalized horizontal tangent vector along the ellipsoidal boundary. 
In this letter, we fix $\gamma=-1$, which corresponds to the extreme case where the spin rate of the fluid is exactly opposite to that of the elliptic deformation.
The geometry is given by a tri-axial ellipsoid with $\beta=0.16$ and $c_*=1$.
This particular case was thoroughly documented both numerically and experimentally in \citet{Grannan:GJI2016}.

\subsubsection{Longitudinal libration}
The ellipsoid boundary is rotating around the $\hat{\mathbf{z}}$-axis with a time varying angular velocity given by 
\begin{linenomath*}
\begin{equation}
\boldsymbol{\Omega}(t) = \left(1+\frac{\Delta\phi\omega_\mathrm{l} }{\Omega_0} \sin(\omega_\mathrm{l} t)\right)\hat{\mathbf{z}},
\end{equation}
\end{linenomath*}
where $\Omega_0$ is the mean rotation rate, $\omega_\mathrm{l}$ the libration frequency and $\Delta \phi$ the libration amplitude.
It is natural to work in the frame of reference attached to the rigid boundary of the ellipsoidal cavity, called librating frame in the following. 
There, the dimensionless inertial force $\mathbf{F}_i$ is 
\begin{linenomath*}
\begin{equation}
\mathbf{F}_i = -2\big[1+\epsilon \sin(ft)\big]\hat{\mathbf{z}}\times \mathbf{u} - \epsilon f \cos(ft)\hat{\mathbf{z}}\times\mathbf{r},
\end{equation}
\end{linenomath*}
where we have introduced the libration frequency $f=\omega_\mathrm{l}/\Omega_0$ and the libration amplitude $\epsilon=\Delta\phi\, f$. We impose no-slip boundary conditions for the velocity field on the fixed boundary of the ellipsoidal cavity.
In this letter, we fix $f=4$, $\epsilon=0.8$ and the geometry is a non-axisymmetric spheroid with $\beta=0.34$ and $c_*=0.81$, which was studied in detail both experimentally by \citet{Grannan:PoF2014} and numerically by \citet{Favier:POF2015}.


\subsubsection{Precession}

The ellipsoid spins at $\boldsymbol{\Omega}(t)=\hat{\mathbf{z}}+P_o\hat{\mathbf{z}}_p(t)$ where $\hat{\mathbf{z}}$ and $\hat{\mathbf{z}}_p(t)$ are the unit vectors along the spin and precession axes, respectively.
$\hat{\mathbf{z}}_p(t)$ is inclined with respect to the vertical $\hat{\mathbf{z}}$ with an angle $\alpha_p$ and is given by
\begin{linenomath*}
\begin{equation}
\hat{\mathbf{z}}_p(t)=\sin(\alpha_p)\cos(t)
\hat{\mathbf{x}}-\sin(\alpha_p)\sin(t)\hat{\mathbf{y}}+
\cos(\alpha_p)\hat{\mathbf{z}}.
\end{equation}
\end{linenomath*}
Following \cite{Tilgner:POF2005,Lin:POF2016}, we work in the reference frame attached to the rigid ellipsoidal cavity, referred to as the mantle frame, and the dimensionless force $\mathbf{F}_i$ acting on the fluid is
\begin{linenomath*}
\begin{equation}
\mathbf{F}_i=-2(\hat{\mathbf{z}}+P_o\hat{\mathbf{z}}_p)\times\mathbf{u}-P_o(\hat{\mathbf{z}}_p \times \hat{\mathbf{z}})\times \mathbf{r},
\end{equation}
\end{linenomath*}
where $P_o=\Omega_p/\Omega_0$ is the Poincar{\'e} number. 
Following the recent dynamo study in spherical geometry of \cite{Lin:POF2016}, we consider the case $\alpha_p=\pi/2$ and $P_o=-0.1$. We impose no-slip boundary conditions for the velocity field.
Knowing that such a flow is dynamo capable in spherical geometry, we here focus on the effect of the ellipticity on the onset of dynamo action in the turbulent regime. 
We therefore consider several cases, from a spherical one with $\beta=0$ and $c_*=1$ to non-axisymmetric ones with $\beta>0$ and $c_*=1/\sqrt{1-\beta^2}$, keeping the fluid volume unchanged as the domain is deformed elliptically.

\subsection{Numerical method and outer insulating domain}

Running numerical simulations in ellipsoidal containers remains a tremendous challenge challenging task.
There have been several attempts using spectral methods for purely hydrodynamical cases \citep[see for example][]{lorenzani2001,SCHMITT2004} but very few results for the magnetohydrodynamical case \citep[see][]{Ivers2017}.
Local methods based on finite elements or finite volumes are more common due to their inherent flexibility for the geometry \citep[see for example][]{Wu:GAFD2009,Cebron:GAFD2012,ernst2013,van2016}, but insulating boundary conditions are then more difficult to consider.
Here, we numerically solve equations~(\ref{Eq:NS})-(\ref{Eq:incom}) using Nek5000, a highly-parallelised spectral-element code developed by Fischer {\it et al.} \citep{FISCHER1997,FISCHER2007,nek5000-web-page}.
Spectral elements can be seen as an intermediate between fully-spectral and local methods.
They combine the geometrical flexibility of finite elements with the exponential convergence of spectral methods.
Nek5000 has been used to study flows driven by tides \citep{favier2014,Grannan:GJI2016,barker2016} and libration \citep{Favier:POF2015}.
Here, we extend these results by also solving for the magnetic potential. All numerical details, as well as several MHD benchmarks and convergence tests, are shown in Supporting Information.

Ideally, one would enclose the fluid cavity in a perfectly insulating outer domain.
This case is usually considered when solving the problem in spherical geometries since the outer potential magnetic field can be implicitly solved using spherical harmonics.
For local methods, this is not possible since the boundary conditions for the magnetic field become non-local.
We choose here to enclose the fluid domain in a larger outer domain with small but finite conductivity \citep{Chan:PEPI2001,Matsui:EPS2004,Guermond:JCP2009}.
The outer surface of the surrounding ellipsoid is given by 
\begin{linenomath*}
\begin{equation}
\frac{x^2}{1+\beta} + \frac{y^2}{1-\beta} +\frac{z^2}{c_*^2} = \chi^2,\label{Eq:Ellip_s}
\end{equation}
\end{linenomath*}
where $\chi>1$ is the aspect ratio between the outer and inner domains.
The velocity field is numerically solved only in the inner domain, whereas the vector potential is solved over the two domains with a jump of magnetic diffusivity defined by a factor $\lambda$ at the interface {(i.e. the CMB)}.
In the insulating outer domain, we thus solve 
\begin{linenomath*}
\begin{equation}
\frac{\partial\mathbf{A}}{\partial t} = {\mathbf{ u_\mathrm{s} \times(\nabla \times A})} +  \lambda\frac{E}{Pm}\nabla^2 \mathbf{A}.\label{Eq:VP_s}
\end{equation}
\end{linenomath*}
For both precession and libration cases, the mantle is not moving so that $\mathbf{u_\mathrm{s}=0}$ in equation~(\ref{Eq:VP_s}). 
For tidally-driven flows however, the outer domain is moving with a constant horizontal tangential velocity so that we cannot neglect the advection term (although it has a negligible contribution since magnetic diffusion is dominant anyway).
{Consistently, the outer magnetic field is advected with a prescribed velocity field given by equation (\ref{eq:uc_tide}), generalised along any elliptical streamline in the outer domain homothetic to the CMB. Note however that this advection term has a negligible contribution since magnetic diffusion is dominant anyway.}

The absence of gauge in (\ref{Eq:VP_f}) and (\ref{Eq:VP_s}) limits us from having a perfectly insulating outer domain surrounding the fluid \citep{Matsui:EPS2004,Cebron:GAFD2012}.
At the interface, all components of the magnetic field and the tangential component of the electric field are continuous \citep{Chan:PEPI2001,Matsui:EPS2004,Matsui:IJCFD2004,Guermond:JCP2009}.
For an infinitely large outer domain, we have asymptotically 
\begin{eqnarray}
\mathbf{B}=O(r^{-3}) \quad \textrm{when} \quad  r\rightarrow \infty \ ,\\
\mathbf{A}=O(r^{-2}) \quad \textrm{when} \quad  r\rightarrow \infty \ .\label{Eq:A_inf}
\end{eqnarray}
For sufficiently large outer domain $\chi\gg1$, the condition (\ref{Eq:A_inf}) can be replaced by 
\begin{linenomath*}
\begin{equation}
\mathbf{A=0} \label{Eq:A_bc}
\end{equation} 
\end{linenomath*}
on the outer boundary \citep{Matsui:IJCFD2004, Chan:PEPI2007}.
Hence we apply the boundary condition (\ref{Eq:A_bc}) on the surface of the boundary given by (\ref{Eq:Ellip_s}).
The initial problem of an infinite insulating outer domain is recovered when both $\lambda\gg1$ and $\chi\gg1$.
A large $\chi$ results in a larger number of grid points and a large $\lambda$ puts  limitations on the convergence properties of our numerical scheme. 
In the Supporting Information, we show that the values $\lambda=100$ and $\chi=8$ used in the following are sufficient to reach convergence \citep[see also][]{Chan:PEPI2001}.

\section{Results and Discussion}

\begin{table}
 \caption{Parameters of all performed simulations, with T, L and P standing for tides, libration and precession, respectively.

Input parameters: Magnetic Prandtl number $Pm$, Ekman number $E$, ellipticity $\beta$, and non-dimensional axial length $c_*$. 

Response parameters: 
 kinetic Reynolds numbers $Re_{z}=U^\mathrm{rms}_{z}/E$ and $Re_\mathrm{tot}=U^\mathrm{rms}/E$, where $U^\mathrm{rms}_{z}=\sqrt{2\langle E_z\rangle}$ and $U^\mathrm{rms}=\sqrt{2\langle E_u\rangle}$, $\langle \cdot \rangle$ representing the time-averaged quantity and $E_u$ the volume-averaged total kinetic energy; and mean growth rate $\sigma$ of the magnetic energy with its standard deviation, determined from the data given in the supporting information. }
 \label{Tab:param}
 \centering
 \begin{tabular}{c c c c c c | c c c c c}
 \hline
 Forcing & Case & $Pm$ & $E\times10^{4}$ & $\beta$ & $c_*$    & $Re_z$ & $Re_\mathrm{tot}$ & $\sigma\times10^{3}$ & Dynamo\\
 \hline
 Tides & T1 & $0.2$ & $2$ & 0.16 & 1.0      & $433$ & 5989 &$-3.04\pm1.5$  &No\\
 Tides &   T2 & $0.5$ & $2$ & 0.16   & 1.0   & $433$ & $5989$ &$1.12\pm0.81$  &Yes\\
 Tides &   T3 & $1.0$ & $2$  & 0.16 & 1.0     & $433$ & $5989$ &$19.6\pm1.6$  &Yes\\
 Tides &   T4 & $2.0$ & $2$ & 0.16  & 1.0     & $433$ & $5989$ &$55.6\pm2.1$  &Yes\\
  \hline
Libration & L1 & $1.0$ & $1.34$ & $0.34$ & 0.81 & $264$ & $2886$ &$-4.2\pm0.41$  &No\\
Libration & L2 & $2.0$ & $1.34$  & $0.34$ &0.81 &  $264$ & $2886$ &$-0.12\pm0.39$  &No\\
Libration & L3 & $3.0$ & $1.34$ & $0.34$ & 0.81  &  $264$ & $2886$ &$2.11\pm0.48$  &Yes\\
Libration & L4 & $4.0$ & $1.34$ & $0.34$  & 0.81 &  $264$ & $2886$ &$5.74\pm0.54$  &Yes\\
  \hline
Precession & P0 & $1.0$ & $1$ & 0.0 & 1.0  &  1138 & $6280$    & $24.1\pm1.1$ &Yes\\
Precession & P1 & $1.0$ & $1$ & 0.1 & 1.005  &  1159 & $6289$   & $36.8\pm1.1$ &Yes\\
Precession & P2 & $1.0$ & $1$ & 0.2 & 1.021  &  1677 & $6239$  & $53.6\pm1.9$ &Yes\\
 \hline
 \end{tabular}
\end{table}

\color{black}

\begin{figure}[ht!]
\centerline{\includegraphics[width=\linewidth]{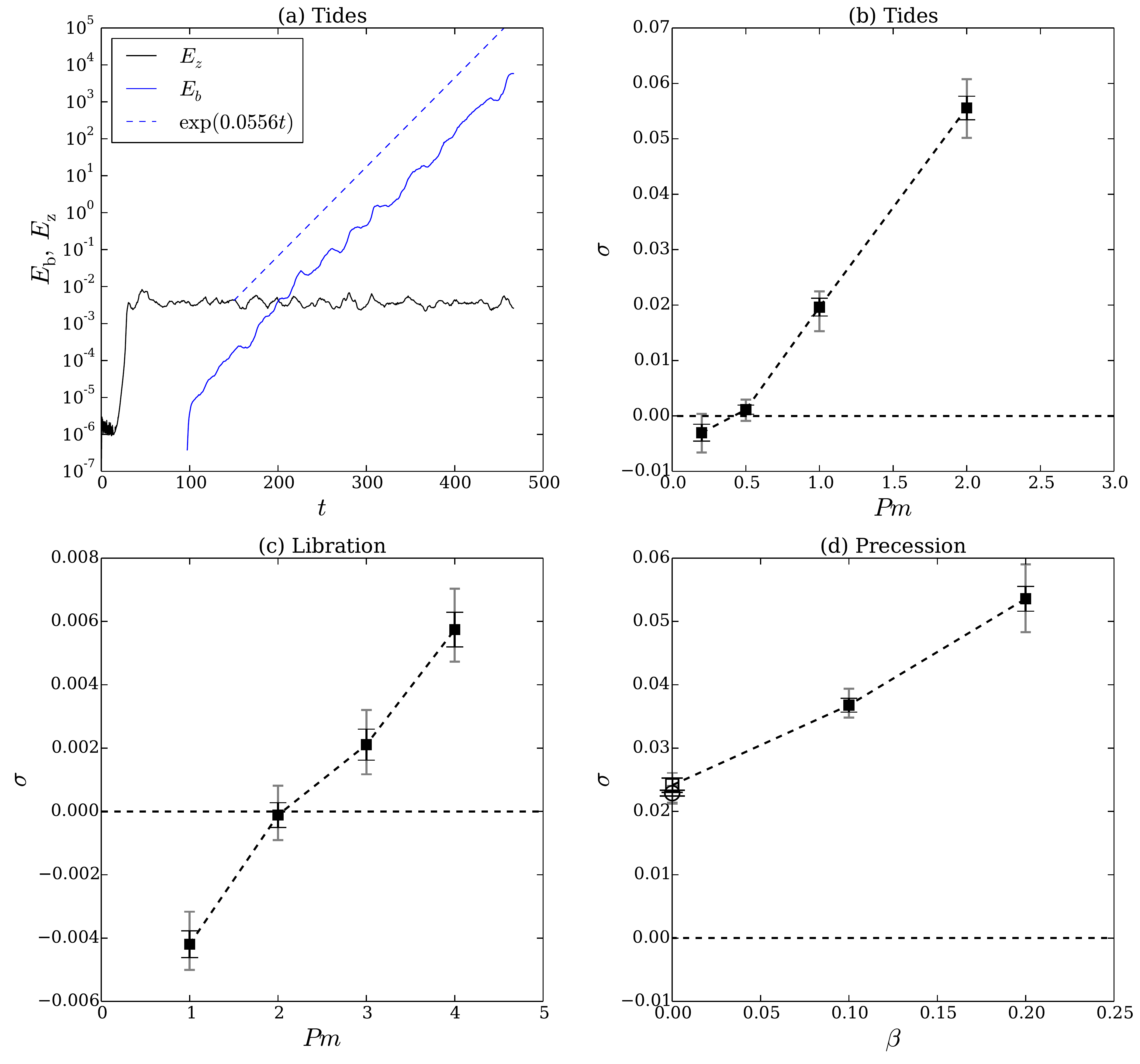}}
\caption{(a) Evolution of the vertical kinetic and magnetic energies of the tidally driven dynamo for $Pm=2$ (case T4, see Table \ref{Tab:param}). (b, c) Growth rate $\sigma$ of the magnetic field for kinematic dynamo driven by tides and libration as a function of the magnetic Prandtl number. (d) growth rate $\sigma$ of the magnetic field for precession dynamo as a function of the equatorial ellipticity $\beta$ of the domain. The hollow circle and square show the mean growth rate in the sphere from \citet{Lin:POF2016} and our case P0, respectively. The black error bars represent one standard deviation and the gray bars represent maximum and minimum estimates.}
\label{Fig:growth}
\end{figure}

For the libration and tides cases considered in this letter, we fix the geometry, forcing amplitude, and Ekman number following previous studies, then we gradually vary the magnetic Prandtl number.
For the precession forcing where dynamo action was already shown in the sphere \citep[][]{Tilgner:POF2005,kida2011turbulent, Lin:POF2016}, we fix the forcing amplitude, Ekman number and magnetic Prandtl number following those previous studies, then we gradually vary the ellipticity to see how it affects the dynamo capability. 
All the parameters are detailed in Table \ref{Tab:param}.
For tides, we consider $E=2\times10^{-4}$ \citep[case $N6$ of][]{Grannan:GJI2016}.
For libration, we choose $E=1.34\times10^{-4}$ \citep[case {$A6$} of][note the different reference length-scale used here]{Favier:POF2015}.
For precession, we choose $E=10^{-4}$ \citep[see Table 1 in][for the spherical equivalent]{Lin:POF2016}.
For all cases, we start with a purely hydrodynamic simulation until it reaches a quasi-steady turbulent state, checked to be consistent with the results previously obtained with a single fluid domain only. 
The details of the mechanism responsible for the turbulent motions have been discussed in details in previous papers \citep[see][and references therein]{le2015flows, Lebars:PRF2016} and we focus here on the magnetohydrodynamic properties of these flows.
We then restart the simulation at a fixed magnetic Prandtl number and solve simultaneously for the magnetic potential including {an initial magnetic perturbation} in the form of an axial dipole in the fluid part of the domain.
This particular magnetic initial condition {(as well as its amplitude)} is irrelevant for the linear kinematic problem and is used here for simplicity.
We have checked that the growth rate of the dynamos presented in this paper does not depend on this arbitrary choice.
 Simulations are pursued until observing exponential growth or decay over many decades (see e.g. Figure~\ref{Fig:growth}(a)). Then, the simulations are restarted using a higher-spectral order and more constraining convergence tolerance in order to check that the results are converged (see more details in the Supporting Information).  

For the first time we obtain self-consistent turbulent kinematic dynamos driven by mechanical forcings in an ellipsoidal geometry.
In Figure~\ref{Fig:growth}(a), we show a typical example corresponding to the tidally-driven case T4 in Table \ref{Tab:param}.
We consider the time evolution of volume-averaged quantities such as the squared vertical velocity component ($E_z$) and magnetic energy ($E_b$) computed using the following definitions
\begin{eqnarray}
E_z(t) &=& \frac{1}{2V}\int_V u_z^2dV,\\
E_b(t) &=& \frac{1}{2V}\int_V \mathbf{B}^2 dV,
\end{eqnarray}
where $V$ is the total volume of the ellipsoidal fluid cavity and $u_z$ is the vertical component of the velocity.
We observe that the magnetic field grows exponentially for $Pm=2$, which clearly shows that our tidally-driven flow is dynamo capable.
Note that the magnetic field growth is non-monotonic.
This is a direct consequence of the very intermittent nature of the underlying turbulent flow, with burst of intense turbulence separated by more quiet phases.
This is characteristic of the saturation of the hydrodynamical instability responsible for the turbulence, the so-called elliptic instability (at least for the tidally-driven and libration-driven cases).
This inevitably leads to significant uncertainties when estimating the growth rate from a finite time series.
We therefore run the simulations for several hundreds of rotation times so that the growth rate can be estimated with a reasonable accuracy.

The main result of this paper is shown in Figure~\ref{Fig:growth}(b, c, d). For  all the mechanical forcings, we obtain kinematic dynamo action.
The critical magnetic Prandtl number is typically around unity for these moderate Reynolds numbers simulations.
Note that the kinetic Reynolds number varies between the different forcings (see Table \ref{Tab:param}) so that it is too early to discuss the relative dynamo efficiency between different cases.
For the precession forcing, we first recover the growth rate obtained in the spherical case ($\beta=0$) by \cite{Lin:POF2016}.
Additionally, we observe that the growth rate of the dynamo increases with $\beta$.
This positive effect of the ellipticity of the dynamo capability of precessional flows could play a major at very low Ekman number where the viscous coupling between the mantle and the fluid becomes less and less important, and only topographic coupling remains {\citep[][]{Jault2007}}.

\begin{figure}
\centerline{\includegraphics[width=\linewidth]{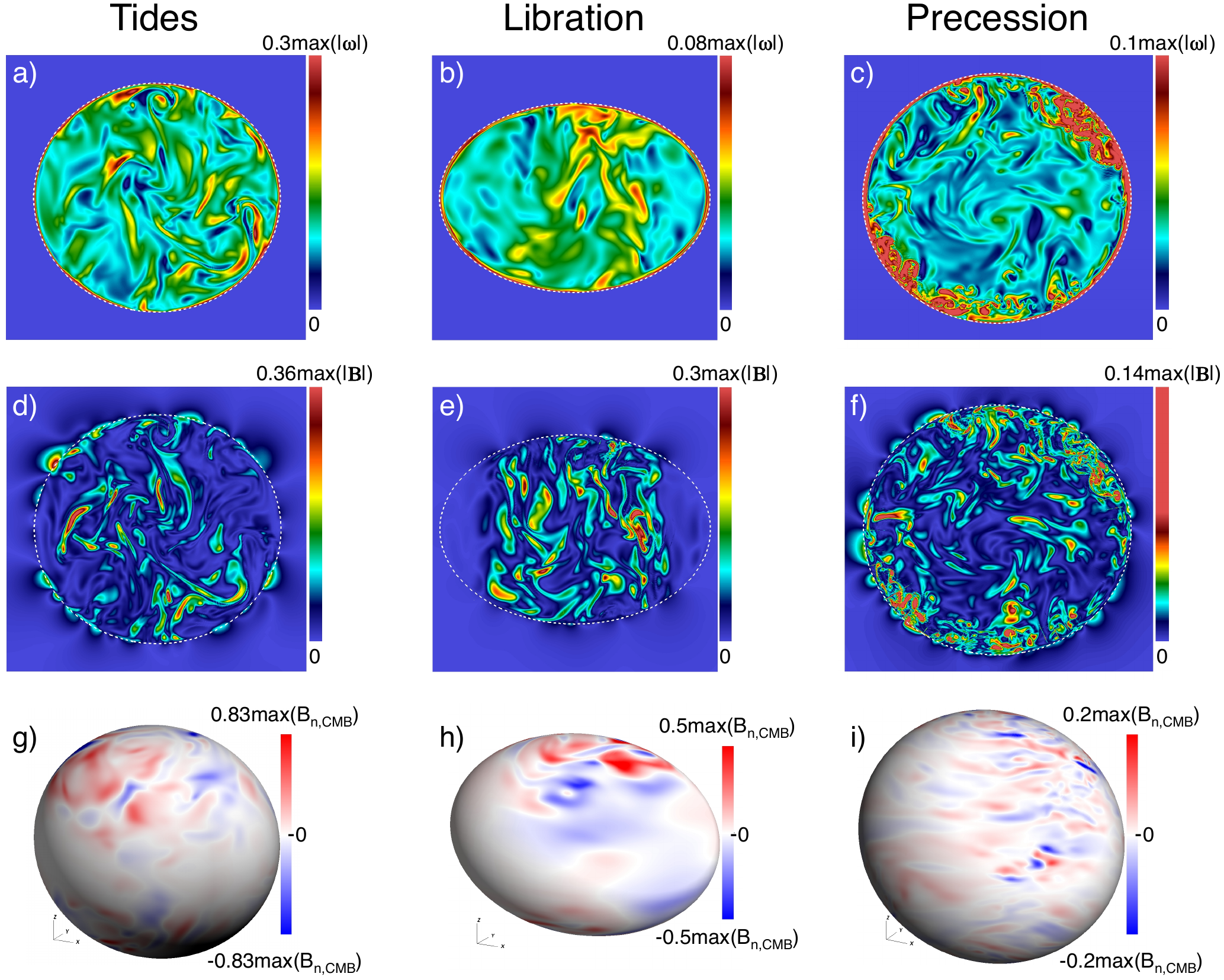}}
\caption{(a, b, c) meridional {(x,z)} slices of the magnitude of the vorticity $|\boldsymbol{\omega} |$, (d, e, f) meridional {(x,z)} slices of the magnitude of  of the magnetic field $|\mathbf{B}|$, and (g, h, i) amplitude of the normal component of the magnetic field at the inner domain boundary (i.e. the CMB) $B_{n,CMB}$, for dynamos driven by tides, libration, and precession in the first, second and third columns, respectively (see cases T4, L4 and P2 in Table \ref{Tab:param}). 
}
\label{Fig:visus}
\end{figure}

Finally, we show in Figure~\ref{Fig:visus} visualizations for each forcing. The magnitudes of the vorticity and magnetic field in a meridional plane, as well as the normal magnetic field at the CMB, are shown. Both tides and libration are very similar, which is not surprising since they rely on the same hydrodynamic elliptical instability mechanism to drive the bulk turbulent flow: vorticity and magnetic field are then spread over the whole domain. Although the kinematic nature of these dynamos does not allow us to perform a detailed analysis of the shape of the magnetic field, we observe, during the exponential growth of the magnetic field and in the absence of any Lorentz force, a dominantly multipolar structure, as also reported for some kinematic precession dynamos in the sphere \citep{tilgner2007kinematic}. For our precession cases, most of the vorticity and magnetic field are concentrated close to the boundary, similarly to the turbulent ring dynamo first shown by \cite{kida2011turbulent} in the sphere: the precessional flow shown here is still strongly related to the viscous coupling between the mantle and the fluid. A bulk instability similar to the two other forcings is, however, expected at lower Ekman number \citep{kerswell1993instability}, which should be explored by additional simulations.

\section{Conclusion}

In conclusion, we have presented the first self-consistent numerical proofs of the existence of mechanically-driven dynamos in geophysically relevant settings, including non-axisymmetric core geometries and fully turbulent flows.
Obviously, as for (even the most recent) convectively-driven dynamos {\cite[see e.g.,][]{schaeffer2017turbulent}}, the dimensionless parameters of our simulations, like the Reynolds and the Ekman numbers, remain far from the planetary ones: turbulence and dynamo thresholds have thus been reached by artificially increasing the amplitude of the mechanical forcings and of the magnetic Prandtl number, respectively.
Also, our results remain up to now limited to kinematic dynamos with no retro-action of the generated magnetic field on the turbulent flow: in this context, no systematic study of the magnetic field topology nor intensity is yet possible.
But we want to stress the following three points.
First, the results presented here already constitute a challenging numerical task, even for the most advanced high performance computing.
Second, the results presented here  aposteriori validate various exotic scenarii of planetary magnetic fields that have been debated during the recent years: e.g., on Mars \citep[][]{arkani2008tidal}, Moon \citep[][]{Lebars2011}, early Earth \citep[][]{andrault2016deep}.
Finally, the results presented here are expected to open new horizons and stimulate additional studies, that we hope will lead to rapid progress in our understanding of those mechanically-driven dynamos. Note in particular that while the models of convectively-driven and mechanically-driven dynamos are often presented as antithetical, they could actually co-exist and even collaborate to produce planetary magnetism \citep[][]{wei2016combined}, a behavior up to now largely unexplored.

Studies of mechanically-driven dynamos are still in their infancy. Meanwhile, it is tempting to already extrapolate on the expected magnetic fields, and more specifically on the possibility of obtaining a large-scale magnetic field of interest for planetary applications. As studied in details by \citet{le2017inertial}, the non-linear saturation of mechanically-driven turbulence can give rise either to large-scale structures or to an inertial wave turbulence. In the former case, the large-scale structures are due to an inverse cascade mechanism from the small scale patterns excited at the instability onset, as classical in rotating turbulence \citep[see also][]{barker2013non, lin2015shear, barker2016turbulence}. Large-scale structures are then capable of sustaining powerful and large-scale magnetic fields \citep[][]{Lin:POF2016}, even if Joule dissipation tends to limit the inverse cascade efficiency \citep[][]{barker2013mag}. The situation is more complex in the later case, where the flow is made of a continuous spectrum of weakly interacting, small-scale inertial waves. But in this case also, the seminal theoretical work of \citet{moffatt1970dynamo} \cite[see also][]{davidson2014dynamics,galtier2014weak} actually predicts the emergence of a large-scale dynamo, which remains to be studied in the context of mechanically-driven flows. Those speculative, but positive conclusions clearly deserve more investigation.

\acknowledgments

This research is funded by the European Research Council (ERC) under the European Unions Horizon 2020 research and innovation program through grant agreement No 681835-FLUDYCO-ERC-2015-CoG. We gratefully acknowledge the computational hours provided on Turing and Ada (Project No. A0020407543) of IDRIS and on HPC resources (Project No. 017B020) of Aix-Marseille Universit\'e financed by the project Equip@Meso (ANR-10-EQPX-29-01) of the program Investissements d'Avenir supervised by the Agence Nationale de la Recherche. We thank Yufeng Lin for providing us with the numerical data of his precessional dynamo in a sphere and David C\'ebron for his comments on an earlier version of this manuscript. 
The spectral element solver Nek5000 is available online \cite[][]{nek5000-web-page}. All numerical results are available upon request from Benjamin Favier (e-mail: favier@irphe.univ-mrs.fr).



\end{document}


%
%


\title{Supporting Information for ``Turbulent kinematic dynamos in ellipsoids driven by mechanical forcing"}
\authors{K. Sandeep Reddy\altaffilmark{1}, Benjamin Favier\altaffilmark{1}, Michael Le Bars\altaffilmark{1}}

\altaffiltext{1}{CNRS, Aix Marseille Univ, Centrale Marseille, IRPHE, Marseille, France}

\begin{article}

\noindent\textbf{Contents of this file}
\begin{enumerate}
\item Implementation within Nek5000
\item Benchmarks
\item Numerical convergence
\end{enumerate}

\vspace{4mm}

\noindent\textbf{Introduction}

In this supplementary document, we provide additional details concerning our numerical approach, and how the magnetic vector potential equation has been implemented within the Nek5000 solver.
Secondly, we benchmark our numerical scheme against classical cases available in the literature.
Thirdly, we discuss the numerical convergence as a function of the values of the numerical parameters introduced by our numerical scheme.

\section{Numerical implementation in Nek5000}

Nek5000 is able to solve MHD problems using the Elsasser variables.
This severely constrains the available boundary conditions for the magnetic field: typically, the  boundary conditions for the velocity and magnetic fields have to be the same.
This is the main reason why we use the magnetic vector potential instead of the magnetic field.
This allows us to use the conjugate solver of Nek5000 where the hydrodynamic flow is solved in a sub-domain while magnetic vector potential equations can be solved in both the inner fluid and outer domains.
Working with the magnetic vector potential also has the advantage of implicitly imposing a divergence-free magnetic field.

We divide the computational domain into $\mathcal{E}$ hexahedral elements.
Within each element, the velocity, vector potential and pressure are represented as the tensor-product Lagrange polynomials of the orders $N$ and $N-2$ based at the Gauss-Lobatto-Legendre and Gauss-Legendre points, respectively.
The degree of freedom scales as $N^3\mathcal{E}$, the numerical convergence is algebraic with increasing the number of elements $\mathcal{E}$ and exponential with increasing the polynomial order $N$.
We use a third-order implicit integration scheme for diffusive terms and a third-order explicit scheme for all remaining nonlinear and inertial terms.
The spectral order we use varies from $N=7$ for tides and libration to $N=9$ for precession (see the numerical convergence tests below) and we use the $3/2$ dealiasing rule to accurately compute the nonlinear terms.
The code is MPI parallelized and we performed our simulations on $1024$ processors typically.

\begin{figure}[h!]
\setfigurenum{S1}
\noindent\includegraphics[width=\linewidth]{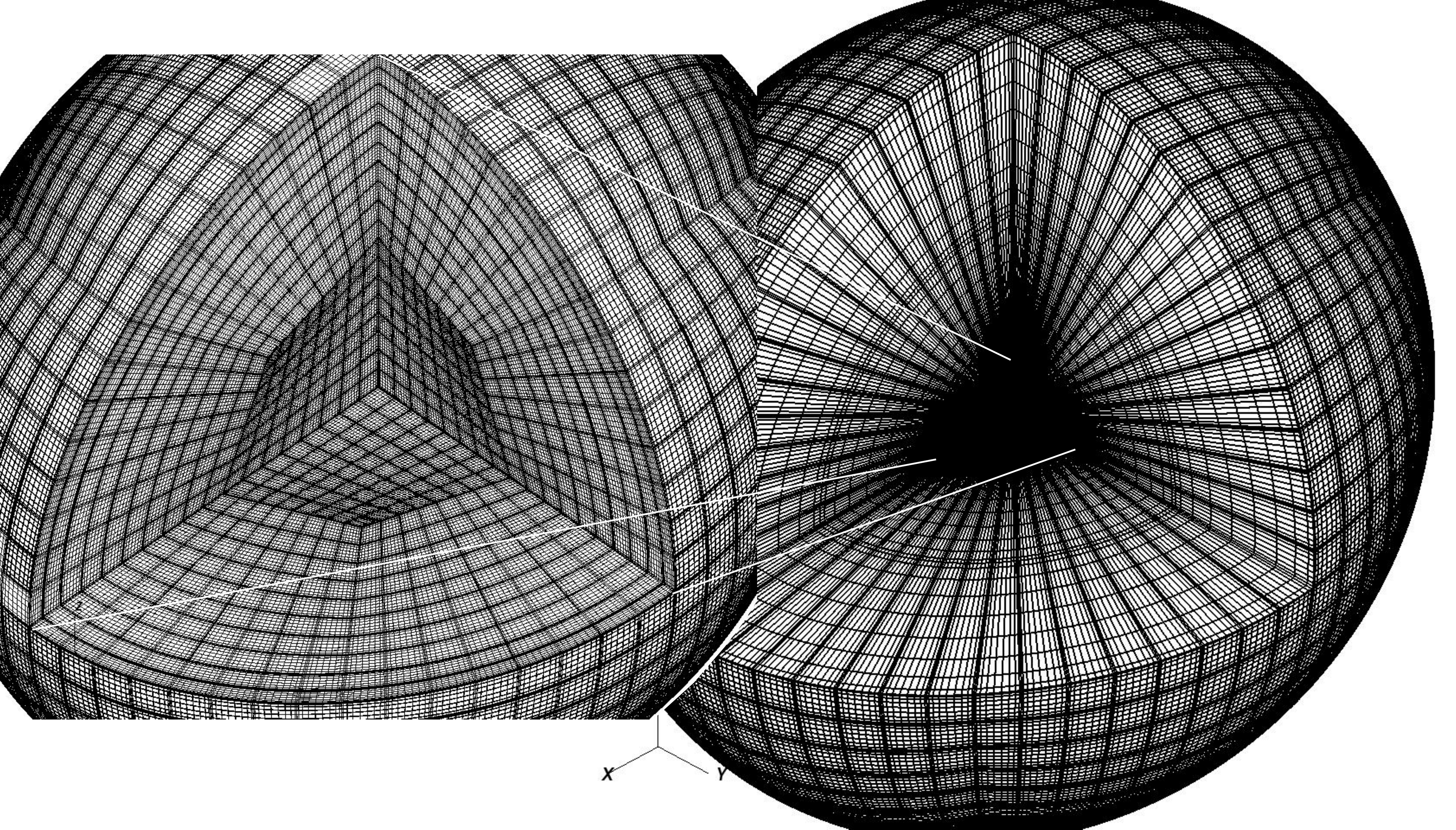}
\caption{Cut through the global ellipsoidal mesh. In the zoomed part we show the mesh for the fluid domain only, which is more densely packed than for the outer domain.}  
\label{fig:mesh}
\end{figure}

In Figure~\ref{fig:mesh} we show the high-resolution mesh divided into 28672 elements. Coupled with a polynomial order $N=9$, this leads to approximately $2\times10^7$ degrees of freedom.
The fluid mesh contains $19456$ elements and is naturally more refined than the much more diffusive outer domain.
To resolve the viscous boundary layers and to resolve the sharp gradients in the magnetic vector potential $\mathbf{A}$ at the interface, we refine the grid on both sides of the boundary of the fluid domain.

\section{Benchmarks}

\subsection{Galloway-Proctor Dynamo}

\begin{figure}[h!]
\setfigurenum{S2}
\noindent\includegraphics[width=\linewidth]{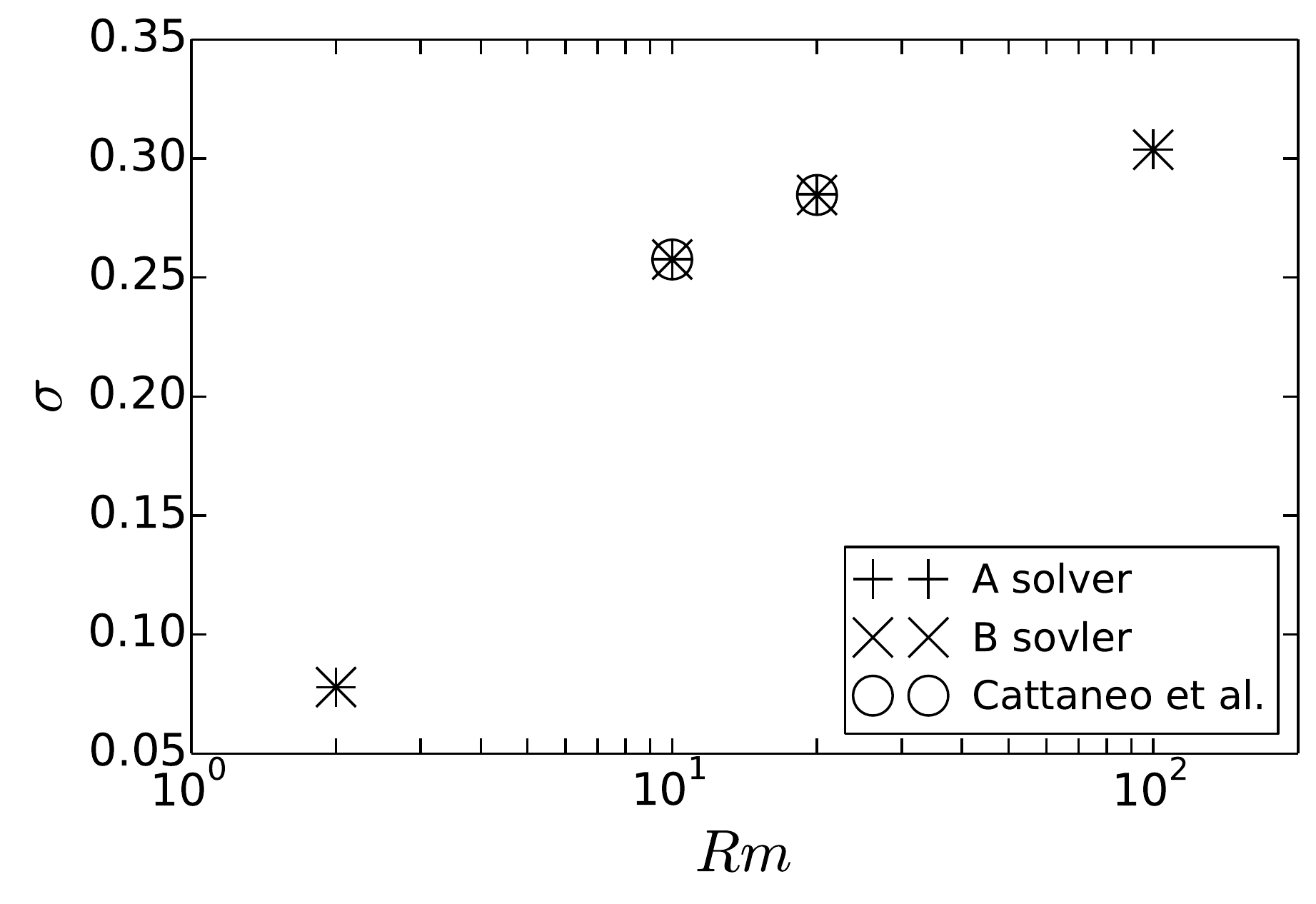}
\caption{Growth rates of the magnetic field for the Galloway-Proctor kinematic dynamo problem versus the magnetic Reynolds number. $+$ correspond to the vector potential method used in this study. Crosses and hollow circles represent the growth rates obtained from the MHD solver of Nek5000 using Elsasser variables and from \citet{Cattaneo:PRL1995}, respectively.}  
\label{fig:GP_growth}
\end{figure}

In order to validate our implementation of the magnetic vector potential equations, we consider the case of the Galloway-Proctor dynamo \citep{Galloway1992,Cattaneo:PRL1995}.
We solve the kinematic dynamo problem in a 3D-periodic Cartesian domain of length $(2\pi,2\pi,2\pi/k_z)$, where $k_z=0.57$ is the optimal wave-number for dynamo action.
The imposed velocity field is given by
\begin{eqnarray}
u_x &=& -\sin(y+\sin\omega t),\\ 
u_y &=&-\cos(x+\cos \omega t),\\
u_z &=&\sin(x+\cos\omega t)+\cos(y+\sin \omega t) \ .
\end{eqnarray}
This kinematic problem is solved both with our magnetic vector potential approach and with the native MHD solver available with Nek5000 based on Elsasser variables. The domain is made of 128 elements and the order of polynomial used within each element is $N=5$. 
Our results are also compared with results obtained using classical pseudo-spectral methods \citep{Cattaneo:PRL1995}.
The initial condition is given by $\mathbf{A}=(0,0,\cos(y) \sin(k_z z))$.
The growth rates for various magnetic Reynolds numbers are shown in Figure~\ref{fig:GP_growth}.
The agreement between the different methods is excellent and validates our implementation of the MHD equations for an idealized periodic case.

\subsection{Freely decaying modes in a conducting sphere}

\begin{figure}[h!]
\setfigurenum{S3}
\noindent\includegraphics[width=\linewidth]{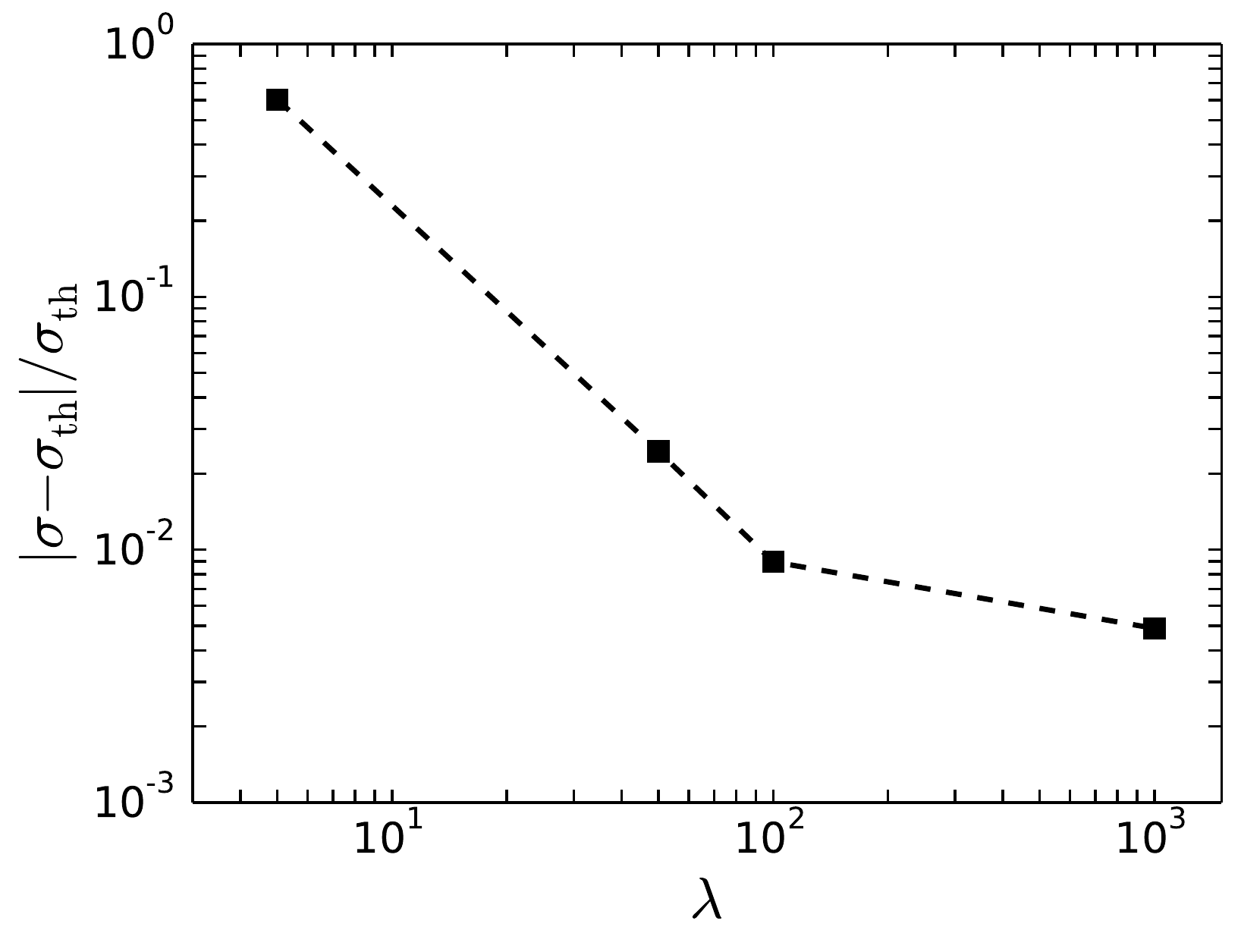}
\caption{Relative error on the decay rate of an axial dipole in a unit sphere as a function of the diffusivity ratio $\lambda$ between the outer and inner domains.}  
\label{fig:decay_sphere}
\end{figure}

In order to check our implementation of the insulating boundary conditions, we consider the purely diffusive decay of a magnetic field inside a conducting sphere of unit radius surrounded by an insulating infinite domain.
We thus solve the simple equation $\partial_t\mathbf{B}=\nabla^2\mathbf{B}$.
This linear equation has an infinite set of eigensolutions. The slowest decay occurs for an axial dipole and is given by \citep[see][]{Iskakov:JCP2004}
\begin{equation}
\label{eq:decay}
{\bf B} = {\bf B}_0\exp(-\pi^2 t) \ .
\end{equation}
Numerically, we solve this problem in a unit sphere surrounded by a larger sphere with radius $\chi=8.5$ and with a magnetic diffusivity increased by a factor $\lambda$. At the outer sphere boundary, we impose $\mathbf{A}=\mathbf{0}$, as in the simulations shown in the paper.
The entire domain is made of 1120 elements, including 448 elements for the inner domain, and the order of polynomial used within each element is $N=5$. 
Our initial condition is an axial dipole of arbitrary amplitude and we consider different diffusivity ratio $\lambda$.
Figure~\ref{fig:decay_sphere} shows the relative error between the measured growth rate $\sigma$ from our simulation and the theoretical prediction $\sigma_\mathrm{th}$ given by equation~(\ref{eq:decay}).
As expected, we observe a rapid decrease of the error as $\lambda$ increases, and the relative error is less than $1\%$ when $\lambda=100$, which is the typical value used throughout this paper.
This test validates our implementation of the insulating boundary conditions using a multi-domain approach with a jump in the magnetic diffusivity at the boundary of the fluid domain.

\subsection{Dudley-James dynamos}

\begin{figure}[h!]
\setfigurenum{S4}
\noindent\includegraphics[width=\linewidth]{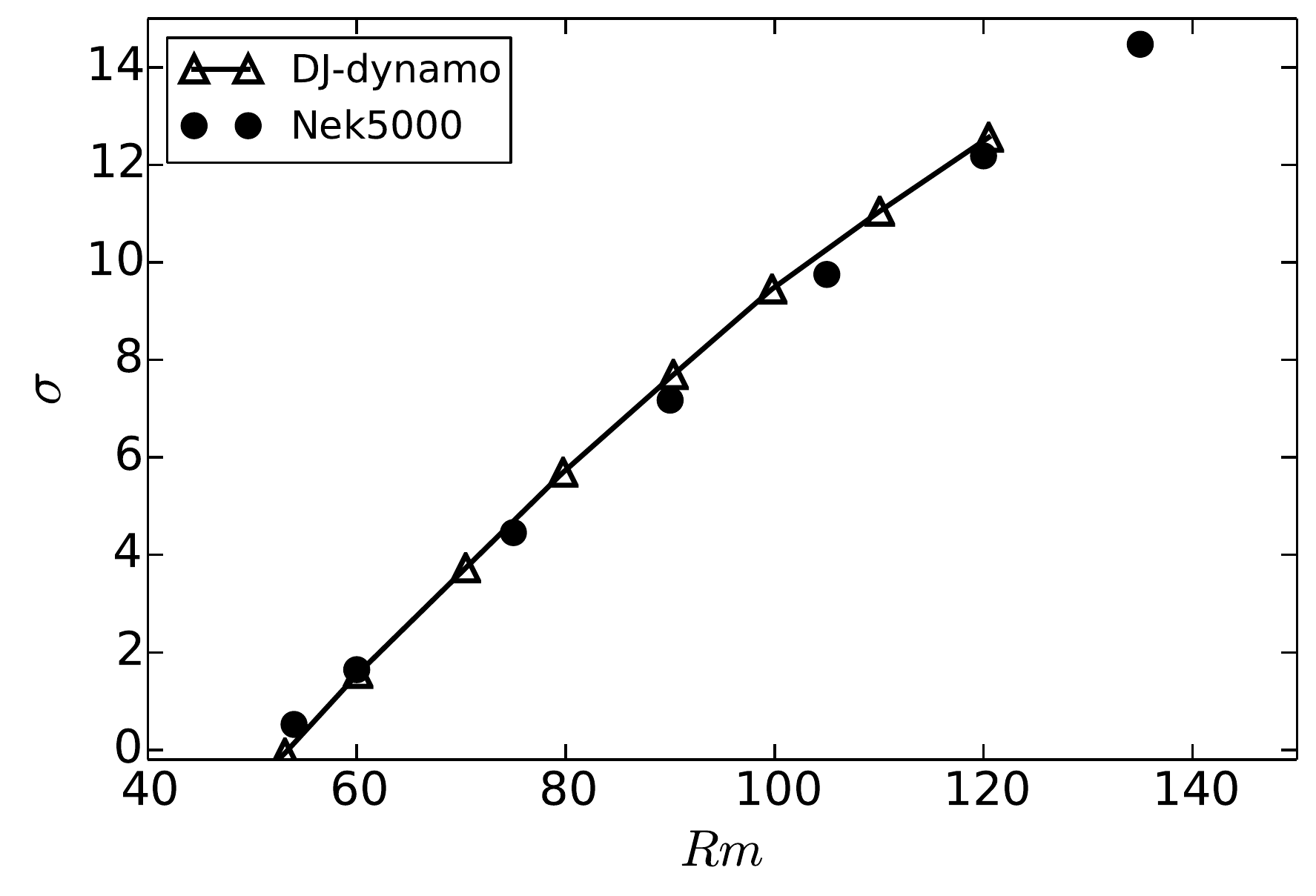}
\caption{Growth rate of the magnetic field for the kinematic dynamo driven by the  $s_2t_2$ flow. Black triangles represent the growth rates extracted from Fig. 8a of \citet{DudleyJames1989}. Black circles correspond to growth rates obtained from our simulations. }  
\label{fig:dj}
\end{figure}

We now consider a test problem when both advection and diffusion of the magnetic field are involved in a confined domain: the so-called Dudley and James dynamos \citep{DudleyJames1989}.
The induction equation is solved in a unit sphere surrounded by an insulating domain.
The imposed flow corresponds to the so-called $s_2t_2$ case of \cite{DudleyJames1989} (see their equation (24) p.421), which is a simple superposition of selected spherical harmonics with prescribed radial structures.
Again, we solve this problem using our numerical approach, fixing $\chi=8.5$, $\lambda=100$ and varying the magnetic diffusivity.
The entire domain is made of 1120 elements, including 448 elements for the inner domain, and the order of polynomial used within each element is $N=5$. 
The initial condition is an equatorial quadrupole of arbitrary amplitude.
We compare the growth rate from the eigenproblem solved by \cite{DudleyJames1989} and those obtained with our initial value numerical integration in Figure~\ref{fig:dj}.
Despite the moderate numerical resolution used, the agreement is excellent, which further confirms that our values of $\chi$ and $\lambda$ are large enough to efficiently mimic an insulating boundary in a kinematic dynamo configuration.

\section{Numerical convergence}

We now discuss how the results shown in this letter depend on our particular choice of numerical and geometric parameters.
We focus on the effect of the spatial resolution through the polynomial order $N$.
The impact of the aspect ratio $\chi$ and diffusivity ratio $\lambda$ are then considered.
In all cases, we consider the particular case $T4$ in Table 1 of the main text, which is a tidally-driven dynamo at $Pm=2$.
This case is particularly demanding due to the large magnetic Prandtl number and due to the fact that the outer domain is actually moving (\textit{i.e.} $\mathbf{u}_s\neq\mathbf{0}$ in equation~(14)) so that large diffusivity ratio is required to accurately model the outer insulating boundary.

\subsection{Convergence with $N$}

\begin{figure}[h!]
\setfigurenum{S5}
\noindent\includegraphics[width=\linewidth]{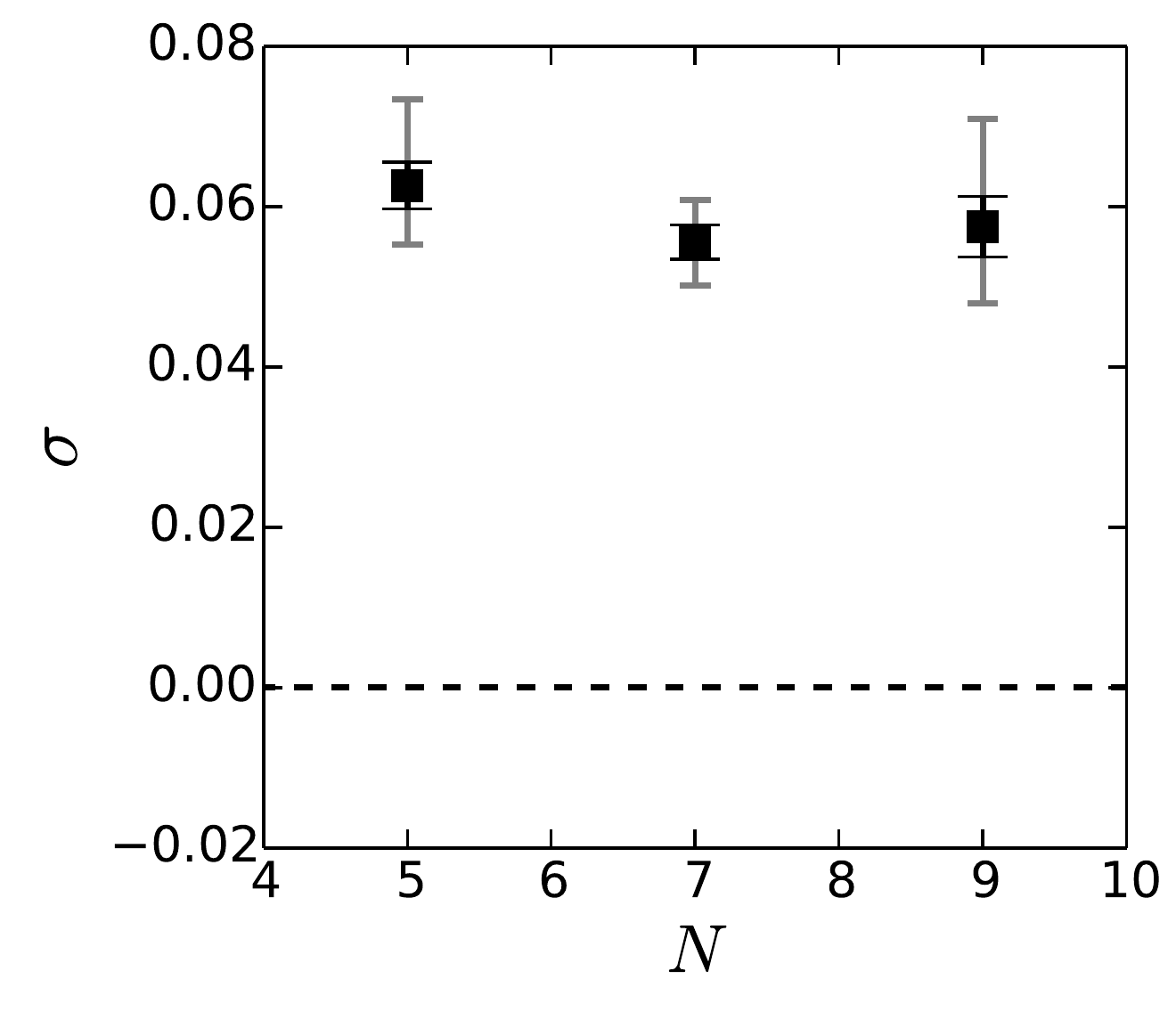}
\caption{Growth rate $\sigma$ of the magnetic energy as a function of the order of polynomial $N$ within each element of the total domain, for the case T4 in Table 1 of the main text. The black error bars represent one standard deviation and the gray bars represent maximum and minimum estimates.}  
\label{fig:Conv_N}
\end{figure}

For case $T4$ in Table 1 of the main text, we performed three simulations which only differ by the polynomial order $N$ to check whether our numerical results are converged. As shown in  Figure~\ref{fig:Conv_N}, 
the resulting growth rate $\sigma$ is nearly the same for $N=7$ and $N=9$.
We conclude that the order of polynomial $N=7$ is sufficient to numerically resolve this problem, and use it for the tidally-driven and libration-driven simulations presented in the main paper.
For the more requiring precession cases, a similar convergence study led us to use $N=9$.

\subsection{Convergence with $\chi$}

\begin{figure}[h!]
\setfigurenum{S6}
\noindent\includegraphics[width=\linewidth]{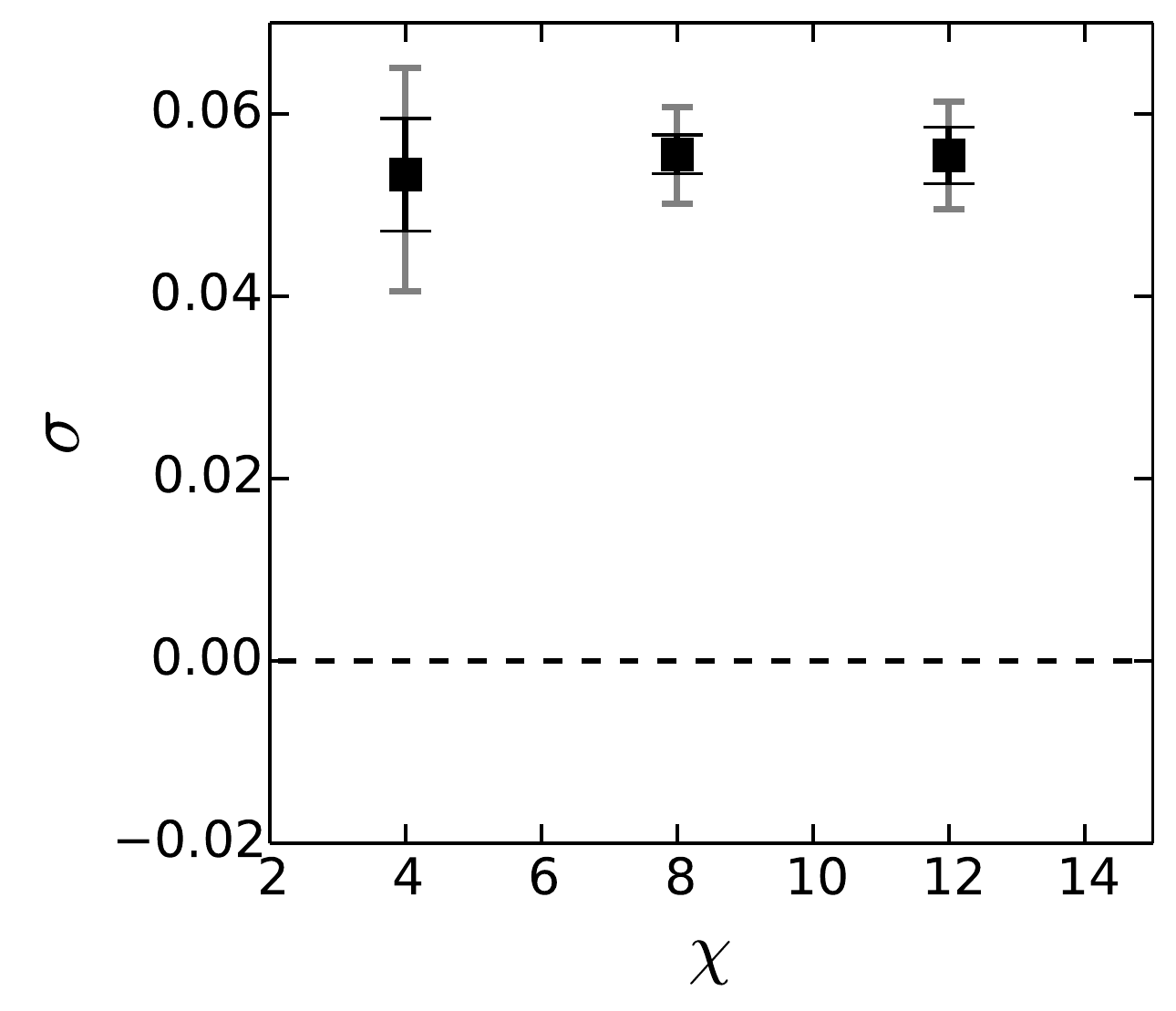}
\caption{Growth rate $\sigma$ of the magnetic energy as a function of the aspect ratio $\chi$ between the outer and the inner domains, for the case T4 in Table 1 of the main text. The black error bars represent one standard deviation and the gray bars represent maximum and minimum estimates.}  
\label{fig:Conv_chi}
\end{figure}

We now consider the effects of changing the aspect ratio $\chi$ on the solution.
We consider different aspect ratios from $\chi=4$ up to $\chi=12$ and compute the growth rate $\sigma$ of the resulting dynamo.
The total number of elements in the mesh for $\chi=4$, $8$ and $12$ are $27136$, $28672$ and $30208$, respectively.
We keep the number of elements in the fluid the same and fix $N=7$ and $\lambda=100$.
In Figure~\ref{fig:Conv_chi}, we show the convergence of $\sigma$ with $\chi$.
While $\chi=4$ slightly underestimates the growth rate, $\chi=8$ and $\chi=12$ are very similar.
We conclude that the choice of $\chi=8$ is sufficient for the artificial effect of the outer boundary to be negligible in our simulations~\citep[see also][]{Chan:PEPI2007}. 

\subsection{Convergence with $\lambda$}

\begin{figure}[h!]
\setfigurenum{S7}
\noindent\includegraphics[width=\linewidth]{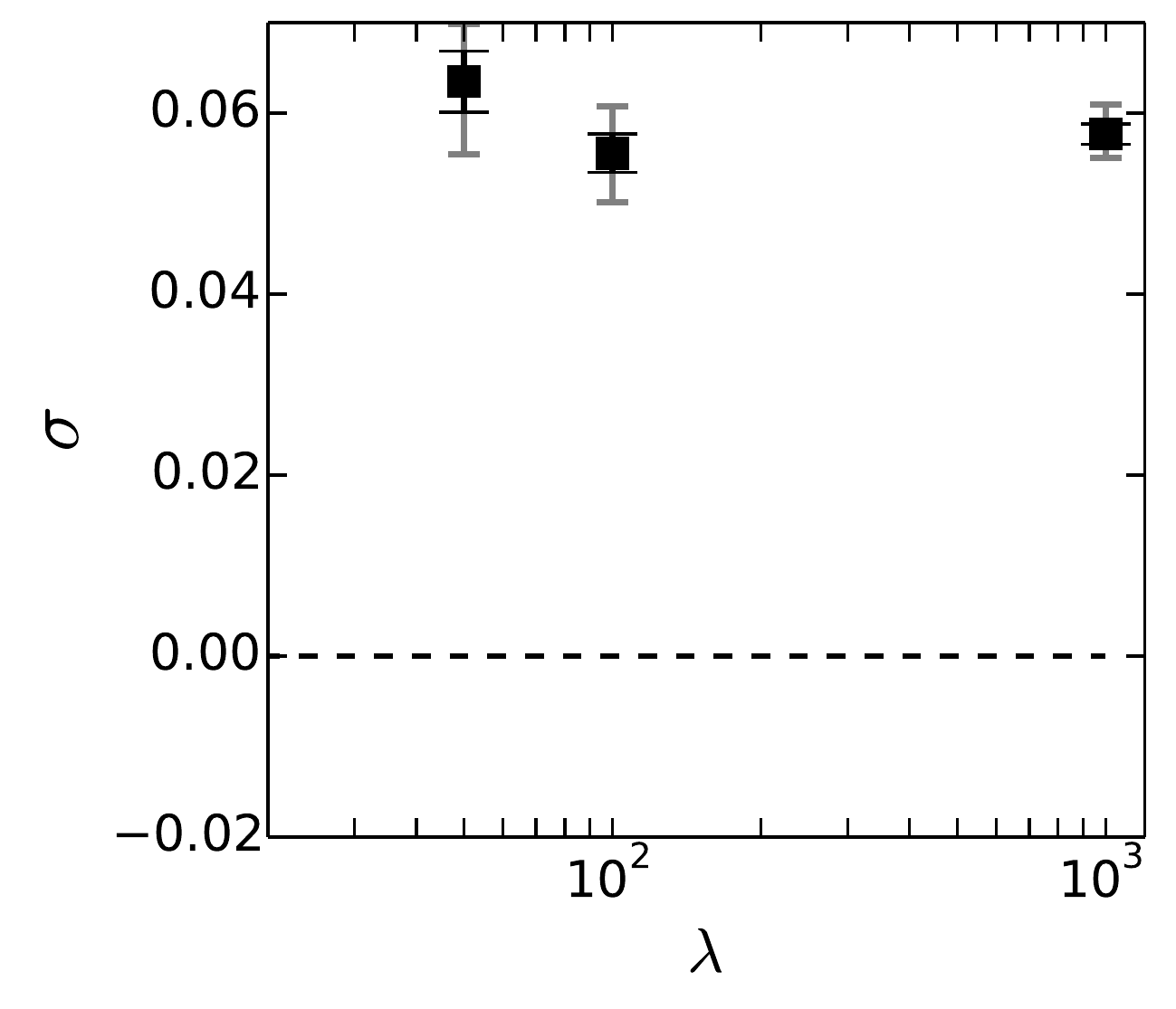}
\caption{Growth rate $\sigma$ of the magnetic energy as a function of the diffusivity ratio $\lambda$ between the outer and the inner domains, for the case T4 in Table 1 of the main text. The black error bars represent one standard deviation and the gray bars represent maximum and minimum estimates.}  
\label{fig:Conv_lambda}
\end{figure}

Finally, let us consider the effect of the ratio of diffusivity between the outer and inner domains.
The cost of a simulation rapidly goes up with increasing $\lambda$ due to the larger number of iterations required to reach a given convergence threshold.
We nevertheless need to choose a large enough ratio $\lambda$ to accurately model the insulating outer domain.
In Figure~\ref{fig:Conv_lambda}, we show the variation of the growth rate $\sigma$, again for case $T4$, for different values of the ratio of diffusivity from $\lambda=50$ up to $\lambda=1000$.
Lower values of $\lambda$ are irrelevant and very demanding numerically since the magnetic field topology is more complex and would require an increase in the resolution used in the outer domain.
We observe that $\lambda=50$ overestimates the growth rate whereas $\lambda=10^2$ is very close to $\lambda=10^3$ with a relative error of less that $2\%$.
We therefore conclude that using $\lambda=10^2$ is enough to accurately model the outer insulating domain.

\end{article}

\bibliographystyle{agufull08}